\documentclass[12pt]{article}
\usepackage{a4wide}
\usepackage{amssymb}
\begin{document}
{\renewcommand{\thefootnote}{\fnsymbol{footnote}}
\hfill  IGPG--07/4--2\\
\begin{center}
{\LARGE  Lattice refining loop quantum cosmology,\\ anisotropic
models and stability}\\
\vspace{1.5em}
Martin Bojowald\footnote{e-mail address: {\tt bojowald@gravity.psu.edu}}
\\
\vspace{0.5em}
Institute for Gravitational Physics and Geometry,
The Pennsylvania State
University,\\
104 Davey Lab, University Park, PA 16802, USA\\
\vspace{1em}
Daniel Cartin\footnote{e-mail address: {\tt Cartin@naps.edu}}
\\
\vspace{0.5em}
Naval Academy Preparatory School, 197 Elliot Street, Newport, RI 02841\\
\vspace{1em}
Gaurav Khanna\footnote{e-mail address: {\tt gkhanna@UMassD.Edu}}
\\
\vspace{0.5em}
Physics Department, University of Massachusetts at Dartmouth,\\ North Dartmouth, MA 02747\\
\vspace{1.5em}
\end{center}
}

\setcounter{footnote}{0}

\newcommand{\case}[2]{{\textstyle \frac{#1}{#2}}}
\newcommand{\lP}{\ell_{\mathrm P}}

\newcommand{\md}{{\mathrm{d}}}
\newcommand{\tr}{\mathop{\mathrm{tr}}}
\newcommand{\sgn}{\mathop{\mathrm{sgn}}}

\newcommand*{\R}{{\mathbb R}}
\newcommand*{\N}{{\mathbb N}}
\newcommand*{\Z}{{\mathbb Z}}
\newcommand*{\Q}{{\mathbb Q}}
\newcommand*{\C}{{\mathbb C}}

\begin{abstract}
 A general class of loop quantizations for anisotropic models is
 introduced and discussed, which enhances loop quantum cosmology by
 relevant features seen in inhomogeneous situations. The main new
 effect is an underlying lattice which is being refined during
 dynamical changes of the volume. In general, this leads to a new
 feature of dynamical difference equations which may not have constant
 step-size, posing new mathematical problems. It is discussed how such
 models can be evaluated and what lattice refinements imply for
 semiclassical behavior. Two detailed examples illustrate that
 stability conditions can put strong constraints on suitable
 refinement models, even in the absence of a fundamental Hamiltonian
 which defines changes of the underlying lattice. Thus, a large class
 of consistency tests of loop quantum gravity becomes available. In
 this context, it will also be seen that quantum corrections due to
 inverse powers of metric components in a constraint are much larger
 than they appeared recently in more special treatments of isotropic,
 free scalar models where they were artificially suppressed.
\end{abstract}

\section{Introduction}

Loop quantum cosmology \cite{LivRev} was designed to test
characteristic effects expected in the full framework of loop quantum
gravity \cite{Rov,ALRev,ThomasRev}. Implementing symmetries at the
kinematical quantum level allows explicit treatments of the dynamical
equations while preserving basic features such as the discreteness of
spatial geometry \cite{SymmRed}. (See also
\cite{SphSymm,AnisoPert,SymmQFT,InhomLattice,Reduction,SymmStatesInt}
for recent work on symmetry reduction in quantum theories.)  Indeed,
several new, initially surprising results were derived in different
applications in cosmology and black hole physics. By now many such
models have been studied in detail.

As the relation of dynamics to that of a possible full framework
without symmetries is not fully worked out, detailed studies can be
used to suggest improvements of the equations for physically viable
behavior. Comparing results with full candidates for quantum dynamics
can then provide stringent self-consistency tests of the overall
framework. It is to be seen if, and how, such alterations of
quantization procedures naturally result from a full quantization. The
first example of this type related to the stability behavior of
solutions to the difference equations of isotropic loop quantum
cosmology, which was studied in \cite{Closed,FundamentalDisc} and was
already restrictive for models with non-zero intrinsic
curvature. Another limitation, realized early on \cite{IsoCosmo},
occurs in the presence of a positive cosmological constant
$\Lambda$. In an exact isotropic model, the extrinsic curvature scale
is given by $k=\dot{a}=\sqrt{8\pi Ga^2\Lambda/3}$ which, due to the
factor of $a^2$, can be large in a late universe although the local
curvature scale $\Lambda$ might be small. Extrinsic curvature plays an
important role since in a flat isotropic model it appears in
holonomies on which loop quantizations are based in such a way that
only $e^{i\alpha k}$ with $\alpha\in{\mathbb R}$ can be represented as
operators, but not $k$ itself \cite{Bohr}.  Large values of $k$ would
either require one to use extremely small $\alpha$ in the relevant
operators, or imply unexpected deviations from classical behavior. In
fact, holonomies as basic objects imply that the Hamiltonian
constraint is quantized to a difference rather than differential
equation \cite{cosmoIV} since $k$ in the Hamiltonian constraint (as in
the Friedmann equation) is not directly quantized but only
exponentials $e^{i\alpha k}$. These are shift operators instead of
differential operators.  For a large, semiclassical universe a
Wheeler--DeWitt wave function should be a good approximation to the
basic difference equation of loop quantum cosmology \cite{SemiClass}
which, in a representation as a function of the momentum $p=a^2$
conjugate to $k$, would be oscillating on scales of the order
$(a\sqrt{\Lambda})^{-1}$. This scale becomes shorter and shorter in an
expanding universe, eventually falling below the discreteness scale of
the difference equation of loop quantum cosmology. At such a point,
discreteness of spatial geometry would become noticeable in the
behavior of the wave function (independently of how physical
observables are computed from it) although the universe should be
classical.

This does not pose a problem for the general formalism, because it
only shows that the specific quantization of the exact isotropic model
used reaches its limits. Physically, this can be understood as a
consequence of a fixed spatial lattice being used throughout the whole
universe evolution. Exponentials $e^{i\alpha k}$ in isotropic models
derive from holonomies $h_e(A)={\cal P}\exp(\int_e
A_a^i\tau_i\dot{e}^a\md t)$ of the Ashtekar connection along spatial
curves $e$. All the freedom contained in choosing edges to capture
independent degrees of freedom of the full theory reduces, in
isotropic models, to the single parameter $\alpha$ which suffices to
separate isotropic connections through all functions $e^{i\alpha
k}$. The parameter $\alpha$, from the full perspective, is thus
related to the edge length used in holonomies. Using a fixed and
constant $\alpha$ is analogous to using only edges of a given
coordinate length, as they occur, for instance, in a regular
lattice. In the presence of a positive cosmological constant, for any
$\alpha$ a value of $k$ will then be reached such that $e^{i\alpha k}$
differs strongly from $i\alpha k$. From the lattice perspective, this
means that the local curvature radius becomes comparable to or smaller
than the fixed lattice scale corresponding to $\alpha$. Such a fixed
lattice ceases to be able to support all small-scale oscillations
relevant for a semiclassical geometry.

This is not problematic if it occurs in a quantum regime where
dynamics is indeed expected to differ from the classical one, but it
poses a problem in semiclassical regimes.
A better treatment has to
refer to changing lattices, which is not easy to implement in a
straightforward quantization of purely homogeneous models.  
In a dynamical equation closer to what is expected from the full
framework, lattice refinements would take place during the evolution
since full Hamiltonian constraint operators generally create new
vertices of a lattice state in addition to changing their edge labels
\cite{RS:Ham,QSDI}. While $k$ increases with increasing volume, the
corresponding $\alpha$ decreases since the lattice is being refined
all the time. For a suitable lattice refinement, the increase in $k$
can be balanced by the decrease of $\alpha$ such that $\alpha k$ stays
small and semiclassical behavior is realized for any macroscopic
volume even with $\Lambda>0$.  This provides an interesting relation
between the fundamental Hamiltonian, which is responsible for the
lattice refinement, and semiclassical properties of models. Testing
whether an appropriate balance between increasing $k$ and lattice
refinements can be reached generically can thus provide stringent
tests on the fundamental dynamics even without using a precise full
Hamiltonian constraint operator.

This feature of lattice refinements was not mimicked in the first
formulations of loop quantum cosmology
\cite{CosmoI,CosmoIII,IsoCosmo,HomCosmo,Bohr} since the main focus was
to understand small-volume effects such as classical singularities
\cite{Sing,BSCG}. In this context, lattice refinements appear irrelevant
because only a few action steps of the Hamiltonian, rather than long
evolution, are sufficient to probe a singularity.  By now,
perturbative regimes around isotropic models have been formulated in
loop quantum cosmology which are inhomogeneous and thus must take into
account lattice states and, at least at an effective level, lattice
refinements \cite{InhomLattice}. One special version, corresponding to
lattices with a number of vertices growing linearly with volume in a
specific way referring to the area operator, has been studied in
detail in isotropic models with a free, massless scalar
\cite{APSII}. Although the complicated relation to a full,
graph-changing Hamiltonian constraint is still not fully formulated,
such models allow crucial tests of the local dynamics.

While isotropic models can easily be understood in terms of wave
functions on a 1-dimensional discrete minisuperspace in terms of
oscillation lengths \cite{DynIn}, anisotropic models with
higher-dimensional minisuperspaces can be more subtle. In such models,
limitations similar to that of a cosmological constant have been
observed as possible instabilities of solutions in classical regions
or the lack of a sufficient number of semiclassical states
\cite{GenFuncBI,InstabLRS,PreClassBI}. For the partial difference
equations of anisotropic models in loop quantum cosmology, stability
issues can be much more severe than in isotropic models and thus lead
to further consistency tests which might help to restrict possible
quantization freedom (see, e.g., \cite{GenFuncKS}). In this paper we
therefore introduce the general setting of anisotropic models taking
into account lattice refinements of Hamiltonian constraint operators,
focusing mainly on the anisotropic model which corresponds to the
Schwarzschild interior. As we will see, the type of difference
equations in general changes since they can become
non-equidistant. This leads to new mathematical problems which we
address here briefly, leaving further analysis for future work. The
examples presented here already show that one can distinguish
different refinement models by their stability properties. The
refinement model corresponding to \cite{APSII} turns out to give
unstable evolution of the Schwarzschild interior, while a new version,
whose vertex number also grows linearly with volume, is stable.
Compared to isotropic models which are sensitive only to how the
vertex number of a state changes with volume, anisotropic models allow
one to test much more detailed properties.

An appendix discusses subtleties in how homogeneous models faithfully
represent inhomogeneous states, mainly regarding the magnitude of
corrections arising from quantizations of inverse metric components
which often plays a large role in cosmological applications.

\section{Difference equation for the Schwarzschild interior with varying discreteness scale}

Basic variables of a loop quantization are holonomies along lattice
links and fluxes over transversal surfaces. For the Schwarzschild
interior \cite{BHInt}, the connection used for holonomies and the
densitized triad used for fluxes take the form
\begin{eqnarray}
A_a^i\tau_i\md x^a &=& \tilde{c}\tau_3\md
x+(\tilde{a}\tau_1+\tilde{b}\tau_2)\md\vartheta
+(-\tilde{b}\tau_1+\tilde{a}\tau_2)\sin\vartheta\md\varphi+
\tau_3 \cos\vartheta \md\varphi \label{sym1}\\
E^a_i\tau^i\frac{\partial}{\partial x^a} &=&
\tilde{p}_c\tau_3\sin\vartheta\frac{\partial}{\partial
x}+(\tilde{p}_a\tau_1+\tilde{p}_b\tau_2)
\sin\vartheta\frac{\partial}{\partial\vartheta}
+(-\tilde{p}_b\tau_1+\tilde{p}_a\tau_2)\frac{\partial}{\partial\varphi}\,.
\label{sym2}
\end{eqnarray}
Coordinates $(x,\vartheta,\varphi)$ are adapted to the symmetry, with
polar angles $\vartheta$ and $\varphi$ along orbits of the rotational
symmetry subgroup, and $\tau_j=-\frac{i}{2}\sigma_j$ in terms of Pauli
matrices.  Spatial geometry is determined by the spatial line element,
which in terms of the densitized triad components is
\begin{equation}\label{metric}
 \md s^2 = \frac{\tilde{p}_a^2+\tilde{p}_b^2}{|\tilde{p}_c|}\md x^2+
 |\tilde{p}_c|\md\Omega^2
\end{equation}
obtained from $q^{ab}=E^a_iE^b_i/|\det E^c_j|$.  We will also
use the co-triad $e_a^i$, i.e.\ the inverse of $e^a_i=E^a_i/\sqrt{|\det
E^b_j|}$,
\begin{equation}\label{cotriadclass} 
e_a^i\tau_i\md x^a = e_c \tau_3 \md x +
(e_a\tau_1 + e_b\tau_2) \md \vartheta + (-e_b \tau_1 +
e_a \tau_2) \sin\vartheta \md\varphi
\end{equation}
with components
\begin{equation}
 e_c = \frac{{\rm sgn}\tilde{p}_c\,
\sqrt{\tilde{p}^2_a+\tilde{p}^2_b}}{\sqrt{|\tilde{p}_c|}}\quad,\quad
 e_b = \frac{\sqrt{|\tilde{p}_c|}\, \tilde{p}_b}
{\sqrt{\tilde{p}^2_a+\tilde{p}^2_b}}\quad {\rm and} \quad
e_a = \frac{\sqrt{|\tilde{p}_c|}\,\tilde{p}_a}
{\sqrt{\tilde{p}^2_a+\tilde{p}^2_b}}\, .
\end{equation}

The phase space is spanned by the spatial constants
$(\tilde{a},\tilde{b},\tilde{c},\tilde{p}_a,\tilde{p}_b,\tilde{p}_c)\in{\mathbb
R}^6$ with non-vanishing Poisson brackets
\[
 \{\tilde{a},\tilde{p}_a\}= \gamma G/L_0\quad,\quad
 \{\tilde{b},\tilde{p}_b\}= \gamma G/L_0 \quad,\quad
 \{\tilde{c},\tilde{p}_c\}= 2\gamma G/L_0
\]
where $G$ is the gravitational constant and $\gamma$ the
Barbero--Immirzi parameter \cite{AshVarReell,Immirzi}. Moreover, $L_0$
is the size of a coordinate box along $x$ used in integrating out the
fields in
\[
 \frac{1}{8\pi\gamma G}\int\md^3 x \dot{A}_a^iE^a_i=
\frac{L_0}{2\gamma G} \dot{\tilde{c}}\tilde{p}_c+ \frac{L_0}{\gamma G}
\dot{\tilde{b}}\tilde{p}_b+ \frac{L_0}{\gamma G}
\dot{\tilde{a}}\tilde{p}_a
\]
to derive the symplectic structure. The SU(2)-gauge transformations
rotating a general triad are partially fixed to U(1) by demanding the
$x$-component of $E^a_i$ to point in the internal $\tau_3$-direction
in (\ref{sym2}). The U(1)-gauge freedom allows one to set
$\tilde{a}=0=\tilde{p}_a$, still leaving a discrete residual gauge
freedom $(\tilde{b},\tilde{p}_b)\mapsto
(-\tilde{b},-\tilde{p}_b)$. The remaining variables can be rescaled by
\begin{equation}
 (b,c):=(\tilde{b},L_o\tilde{c})\quad,\quad
 (p_b,p_c):=(L_o\tilde{p}_b,\tilde{p}_c)\,.
\end{equation}
to make the canonical structure $L_0$-independent:
\begin{equation}
 \{b,p_b\}=\gamma G\quad,\quad \{c,p_c\}= 2\gamma G\,.
\end{equation}
This rescaling is suggested naturally by holonomies, as written below,
and fluxes which are considered the basic objects in loop
quantizations.

To express the elementary variables through holonomies, which unlike
connection components will be promoted to operators, it suffices to
choose curves along the $x$-direction of coordinate length $\tau L_0$
and along $\vartheta$ of coordinate length $\mu$ since this captures
all information in the two connection components,
\begin{eqnarray}
 h^{(\tau)}_x(A) &=& \exp\int_0^{\tau L_o}\md x \tilde{c}\tau_3=
\cos\frac{\tau c}{2}+ 2\tau_3\sin\frac{\tau c}{2} \label{hol1}\\
h^{(\mu)}_{\vartheta}(A) &=& \exp\int_0^{\mu}\md\vartheta \tilde{b}\tau_2=
\cos\frac{\mu b}{2}+ 2\tau_2\sin\frac{\mu b}{2}\, .\label{hol3}
\end{eqnarray}
The quantum Hilbert space is then based on cylindrical states
depending on the connection through countably many holonomies, which
can always be written as almost periodic functions $f(b,c) =
\sum_{\mu,\tau}\, f_{\mu,\tau} \exp {\frac{i}{2}\, (\mu b + \tau c)}$ of
two variables. These form the set of functions on the double product
of the Bohr compactification of the real line, which is a compact
Abelian group. Its Haar measure defines the inner product of the
(non-separable) Hilbert space, in which states
\begin{equation} \label{basis}
 \langle b,c|\mu,\tau\rangle = e^{\frac{i}{2}\,(\mu b+\tau c)} \qquad
 \mu,\tau\in{\R}\, .
\end{equation}
form an orthonormal basis. Holonomies simply act by multiplication on
these states, while densitized triad components become derivative
operators
\begin{equation}
 \hat{p}_b = -i{\gamma\lP^2}\,\frac{\partial}{\partial b},
\quad\quad \hat{p}_c = -2i\gamma\lP^2 \frac{\partial}{\partial c}
\end{equation}
using the Planck length $\lP=\sqrt{G\hbar}$.  They act as
\begin{equation}
 \hat{p}_b|\mu,\tau\rangle =
\textstyle{\frac{1}{2}}\,\gamma\lP^2\, \mu |\mu,\tau\rangle,\qquad
\hat{p}_c|\mu,\tau\rangle = \gamma\lP^2\, \tau
 |\mu,\tau\rangle\, ,
\end{equation}
immediately showing their eigenvalues.

To formulate the dynamical equation, one has to quantize the
Hamiltonian constraint
\begin{equation} \label{ham} 
H = \frac{1}{\gamma^2}\, \int \md^3 x\,\,
\epsilon_{ijk} (- \underline{F}_{ab}^k+\gamma^2\, \Omega_{ab}^k)
\frac{E^{ai}E^{bj}}{\sqrt{|\det E|}}
\end{equation}
where $\Omega_{ab}^k\tau_k\md x^a\wedge\md
x^b=-\sin\vartheta\tau_3\md\vartheta\wedge\md\varphi$ is the intrinsic
curvature of 2-spheres, while $\underline{F}_{ab}^k$ is the curvature
computed from $A_a^i$ ignoring the spin connection term
$\sin\vartheta\tau_3\md\varphi$. Following standard procedures a
Hamiltonian constraint operator can be expressed in the basic
operators. First, one replaces the inverse determinant of $E^a_i$ by a
Poisson bracket, following \cite{QSDI},
\begin{equation}\label{cotriad}
\epsilon_{ijk}\tau^i\frac{E^{aj}E^{bk}}{\sqrt{|\det E|}}=
-\frac{1}{4\pi\gamma G}
\sum_{K\in\{x,\vartheta,\varphi\}} \frac{1}{\ell_0^K}\epsilon^{abc}\omega_c^K
h_K^{(\delta)}\{h_K^{(\delta)-1},V\}
\end{equation}
with edge lengths $\ell_0^x=\delta L_0$ and
$\ell_0^{\vartheta/\varphi}=\delta$, and left-invariant 1-forms
$\omega_c^K$ on the symmetry group manifold.  For curvature components
$\underline{F}_{ab}^k$ one uses a holonomy around a closed loop
\begin{equation} \label{F}
\underline{F}_{ab}^i(x)\tau_i=\frac{\omega^I_a\omega^J_b}{{\cal A}_{(IJ)}}
(h^{(\delta)}_{IJ}-1)\, +\,\,O((b^2+c^2)^{3/2}\sqrt{\cal A})
\end{equation}
with
\begin{equation} \label{loop}
 h^{(\delta)}_{IJ} = h_I^{(\delta)}h_J^{(\delta)}(h_I^{(\delta)})^{-1}
(h_J^{(\delta)})^{-1}
\end{equation}
and ${\cal A}_{IJ}$ being the coordinate area of the loop, using the
corresponding combinations of $\ell_0^I$. In these expressions, a
parameter $\delta$ has been chosen which specifies the length of edges
with respect to the background geometry provided by the symmetry
group. Putting all factors together and replacing Poisson brackets by
commutators, one has
\begin{eqnarray}
 \hat{H}^{(\delta)} &=& 2i
(\gamma^3\delta^3\lP^2)^{-1} \tr\left(\sum_{IJK}
\epsilon^{IJK}\hat{h}_I^{(\delta)} \hat{h}_J^{(\delta)} \hat{h}_I^{(\delta)-1}
\hat{h}_J^{(\delta)-1}\hat{h}_K^{(\delta)}
[\hat{h}_K^{(\delta)-1},\hat{V}]+2\gamma^2\delta^2\tau_3 \hat{h}_x^{(\delta)}
[\hat{h}_x^{(\delta)-1},\hat{V}]\right)\nonumber\\\nonumber &=&
4i(\gamma^3\delta^3\lP^2)^{-1}\left(
 8\sin\frac{\delta b}{2}\cos\frac{\delta b}{2} \sin\frac{\delta
c}{2}\cos\frac{\delta c}{2}
 \left(\sin\frac{\delta b}{2}\hat{V}\cos\frac{\delta b}{2}-
 \cos\frac{\delta b}{2}\hat{V}\sin\frac{\delta b}{2}\right)\right.\\
 && + \left.\left(4\sin^2\frac{\delta b}{2}\cos^2\frac{\delta
b}{2}+\gamma^2\delta^2\right)
 \left(\sin\frac{\delta c}{2}\hat{V}\cos\frac{\delta c}{2}-
 \cos\frac{\delta c}{2}\hat{V}\sin\frac{\delta c}{2}\right)\right)\label{C}
\end{eqnarray}
which acts as
\begin{eqnarray}
 \hat{H}^{(\delta)} |\mu,\tau\rangle &=&
(2\gamma^3\delta^3\lP^2)^{-1}
\left[ 2(V_{\mu+\delta,\tau}-V_{\mu-\delta,\tau})\right.\\
&&\times(|\mu+2\delta,\tau+2\delta\rangle- |\mu+2\delta,\tau-2\delta\rangle-
|\mu-2\delta,\tau+2\delta\rangle+ |\mu-2\delta,\tau-2\delta\rangle)\nonumber\\
&&+\left.(V_{\mu,\tau+\delta}-V_{\mu,\tau-\delta}) (|\mu+4\delta,\tau\rangle-
2(1+2\gamma^2\delta^2)|\mu,\tau\rangle+ |\mu-4\delta,\tau\rangle)\right]
\nonumber
\end{eqnarray}
on basis states.  This operator can be ordered symmetrically, defining
$\hat{H}_{\rm symm}^{(\delta)}:=\frac{1}{2} (\hat{H}^{(\delta)}+
\hat{H}^{(\delta)\dagger})$, whose action is\footnote{Note that the
first factor of 2 in the next-to-last line was missing in
\cite{BHInt} and analogous places in subsequent formulas. 
This turns out to be crucial for the stability analysis below.
In particular, with the corrected coefficient the quantization of the
Schwarzschild interior in \cite{BHInt} is unstable for all values of
$\gamma$. Possible restrictions on $\gamma$, as suggested in
\cite{GenFuncKS} based on a difference equation with the wrong
coefficient, then do not follow easily but could be obtained from a
more detailed analysis.}
\begin{eqnarray}
 \hat{H}_{\rm symm}^{(\delta)} |\mu,\tau\rangle &=&
(2 \gamma^3\delta^3\lP^2)^{-1}\,
\left[(V_{\mu+\delta,\tau}-V_{\mu-\delta,\tau}+V_{\mu+3\delta,\tau+2\delta}-
V_{\mu+\delta,\tau+2\delta}) |\mu+2\delta,\tau+2\delta\rangle \right.\nonumber\\
&&- (V_{\mu+\delta,\tau}-V_{\mu-\delta,\tau}+V_{\mu+3\delta,\tau-2\delta}-
V_{\mu+\delta,\tau-2\delta}) |\mu+2\delta,\tau-2\delta\rangle \nonumber\\
&&-(V_{\mu+\delta,\tau}-V_{\mu-\delta,\tau}+V_{\mu-\delta,\tau+2\delta}-
V_{\mu-3\delta,\tau+2\delta}) |\mu-2\delta,\tau+2\delta\rangle \nonumber\\
&&+ (V_{\mu+\delta,\tau}-V_{\mu-\delta,\tau}+V_{\mu-\delta,\tau-2\delta}-
V_{\mu-3\delta,\tau-2\delta}) |\mu-2\delta,\tau-2\delta\rangle \nonumber\\
&&+{\textstyle\frac{1}{2}}(V_{\mu,\tau+\delta}-V_{\mu,\tau-\delta}+V_{\mu+4\delta,\tau+\delta}-
V_{\mu+4\delta,\tau-\delta}) |\mu+4\delta,\tau\rangle \nonumber\\
&&-2(1+2\gamma^2\delta^2)(V_{\mu,\tau+\delta}-V_{\mu,\tau-\delta}) |\mu,\tau\rangle
\nonumber\\
&&+\left.{\textstyle\frac{1}{2}}
(V_{\mu,\tau+\delta}-V_{\mu,\tau-\delta}+V_{\mu-4\delta,\tau+\delta}-
V_{\mu-4\delta,\tau-\delta}) |\mu-4\delta,\tau\rangle\right]\,.
\end{eqnarray}
Transforming this operator to the triad representation obtained as
coefficients of a wave function $|\psi\rangle = \sum_{\mu,\tau} \,
\psi_{\mu,\tau}|\mu,\tau\rangle$ in the triad eigenbasis and using the
volume eigenvalues
\[
 V_{\mu,\tau}= 4\pi \sqrt{|(\hat{p}_c)_{\mu,\tau}|}
(\hat{p}_b)_{\mu,\tau}= 2\pi (\gamma\lP^2)^{3/2} \sqrt{|\tau|}\mu\,,
\]
a difference equation
\begin{eqnarray} \label{qee2}
&&\frac{\gamma^{3/2}\delta^3}{\pi\lP}
(\hat{H}_{\rm symm}^{(\delta)}|\psi\rangle)_{\mu,\tau}
=2\delta(\sqrt{|\tau+2\delta|}+\sqrt{|\tau|})
\left(\psi_{\mu+2\delta,\tau+2\delta}- \psi_{\mu-2\delta,\tau+2\delta}\right)
\nonumber\\
&& \qquad+(\sqrt{|\tau+\delta|}-\sqrt{|\tau-\delta|})
\left((\mu+2\delta)\psi_{\mu+4\delta,\tau}-
2(1+2\gamma^2\delta^2)\mu\psi_{\mu,\tau}+
(\mu-2\delta)\psi_{\mu-4\delta,\tau}\right)\nonumber\\
&&\qquad+2\delta(\sqrt{|\tau-2\delta|}+\sqrt{|\tau|})
\left(\psi_{\mu-2\delta,\tau-2\delta}-
\psi_{\mu+2\delta,\tau-2\delta}\right)\nonumber\\
&=& 0
\end{eqnarray}
results for physical states. (For small $\mu$ the equation has to be
specialized further due to the remaining gauge freedom; see
\cite{BHInt}. This is not relevant for our purposes.)

\subsection{Relation to fixed lattices}

Although there are no spatial lattices appearing in the exactly
homogeneous context followed here, the construction of the Hamiltonian
constraint mimics that of the full theory. States are then associated
with spatial lattices, and holonomies refer to embedded edges and
loops.  The parameter $\delta$ is the remnant of the loop size (in
coordinates) used to act with holonomies on a spatial lattice. As one
can see, this parameter is important for the resulting difference
equation, determining its step-size. The above construction, using a
constant $\delta$, can be seen as corresponding to a lattice chosen
once and for all such that the loop size is not being adjusted even
while the total volume increases. As described in the introduction,
this ignores the possible creation of new lattice vertices and links,
and can be too rigid in certain semiclassical regimes.

To express this clearly, we now construct holonomies which are not
simply along a single edge of a certain length $\delta$, but which are
understood as holonomies along lattice links. We keep our coordinate
box of size $L_0$ in the $x$-direction as well as the edge length
$\ell_0$. If this is a link in a uniform lattice, there are
${\cal N}_x=L_0/\ell_0$ lattice links in this direction,
and a link holonomy appears in the form
\begin{equation}
 h_x=\exp(\ell_0\tilde{c}\tau_3)= \exp(\ell_0 c\tau_3/L_0)=
\exp(c\tau_3/{\cal N}_x)
\end{equation}
when computed along whole lattice edges.  Thus, a constant coefficient
$1/{\cal N}_x$ in holonomies corresponds to a fixed lattice whose
number of vertices does not change when the volume increases. Lattice
refinements of an inhomogeneous lattice state, on the other hand, can
be mimicked by a parameter ${\cal N}_x$ which depends on the phase
space variables, most importantly the triad components. If this is
carried through, as we will see explicitly below, the step-size of the
resulting difference equation is not constant in the triad variables
anymore.

\subsection{Lattice refinements}

Let us now assume that we have a lattice with ${\cal N}$ vertices in a
form adapted to the symmetry, i.e.\ there are ${\cal N}_x$ vertices
along the $x$-direction (whose triad component $p_c$ gives rise to the
label $\tau$) and ${\cal N}_{\vartheta}^2$ vertices in spherical
orbits of the symmetry group (whose triad component $p_b$ gives rise
to the label $\mu$). Thus, ${\cal N}={\cal N}_x{\cal
N}_{\vartheta}^2$.

Since holonomies in such a lattice setting are computed along single
links, rather than through all of space (or the whole cell of size
$L_0$), basic ones are $h_x=\exp(\ell_0^x\tilde{c}\tau_3)$ and
$h_{\vartheta}=\exp(\ell_0^{\vartheta}\tilde{b}\tau_2)$, denoting the
edge lengths by $\ell_0^I$ and keeping them independent of each other
in this anisotropic setting. Edge lengths are related to the number of
vertices in each direction by $\ell_0^x=L_0/{\cal N}_x$ and
$\ell_0^{\vartheta}=1/{\cal N}_{\vartheta}$. With the rescaled
connection components $c=L_0\tilde{c}$ and $b=\tilde{b}$ we have basic
holonomies
\begin{equation}
 h_x=\exp(\ell_0^xL_0^{-1} c\tau_3)=\exp(c\tau_3/{\cal N}_x)\quad,\quad
 h_{\vartheta}= \exp(\ell_0^{\vartheta}b\tau_2)=
 \exp(b\tau_2/{\cal N}_{\vartheta})\,.
\end{equation}
Using this in the Hamiltonian constraint operator then gives a
difference equation whose step-sizes are $1/{\cal N}_I$. 

So far, we only reinterpreted $\delta$ in terms of vertex numbers. We
now turn our attention to solutions to the Hamiltonian constraint
which, in the full theory, usually changes the lattice by adding new
edges and vertices while triad eigenvalues increase. For larger $\mu$
and $\tau$, the Hamiltonian constraint thus acts on a finer lattice
than for small values, and the parameter ${\cal N}$ for holonomies
appearing in the constraint operator is not constant on phase space
but triad dependent. Due to the irregular nature of lattices with
newly created vertices such a refinement procedure is difficult to
construct explicitly. But it is already insightful to use an effective
implementation, using the derivation of the Hamiltonian constraint for
a fixed lattice, but assuming the vertex number ${\cal N}(\mu,\tau)$
to be phase space dependent. Moreover, we include a parameter $\delta$
as before, which now takes a value $0<\delta<1$ and arises because a
graph changing Hamiltonian does not use whole lattice edges but only a
fraction, given by $\delta$.\footnote{A precise value can be
determined only if a precise implementation of the symmetry for a
fixed full constraint operator is developed. Currently, both the
symmetry reduction for composite operators and a unique full
constraint operator are lacking to complete this program and we have
to work with $\delta$ as a free parameter. This parameter is sometimes
related to the lowest non-zero eigenvalue of the full area operator
\cite{Bohr,APSII}. From the inhomogeneous perspective of lattice
states used here, however, there is no indication for such a
relation.}  Effectively assuming in this way that
the lattice size is growing through the basic action of the
Hamiltonian constraint, we will obtain a difference equation whose
step-size $\delta/{\cal N}$ is not constant in the original triad
variables.

For the Schwarzschild interior, we have step sizes $\delta/{\cal
N}_{\vartheta}$ for $\mu$ and $\delta/{\cal N}_x$ for $\tau$.  Going
through the same procedure as before, we end up with an operator
containing flux-dependent holonomies instead of basic ones, e.g.,
${\cal N}_x(\mu,\tau)h_x={\cal N}_x(\mu,\tau)\exp(c\tau_3/{\cal
N}_x(\mu,\tau))$ which reduces to an ${\cal N}_x$-independent
connection component $c$ in regimes where curvature is small. Keeping
track of all prefactors and holonomies in the commutator as well as
the closed loop, one obtains the difference equation
\begin{eqnarray}
&&C_+(\mu,\tau)
\left(\psi_{\mu+2\delta {\cal N}_{\vartheta}(\mu,\tau)^{-1},
\tau+2\delta {\cal N}_x(\mu,\tau)^{-1}}- 
\psi_{\mu-2\delta {\cal N}_{\vartheta}(\mu,\tau)^{-1},
\tau+2\delta {\cal N}_x(\mu,\tau)^{-1}}\right)
\nonumber\\
&& + C_0(\mu,\tau)
\left((\mu+2\delta {\cal N}_{\vartheta}(\mu,\tau)^{-1})
\psi_{\mu+4\delta {\cal N}_{\vartheta}(\mu,\tau)^{-1},\tau}-
2(1+2\gamma^2\delta^2 {\cal N}_{\vartheta}(\mu,\tau)^{-2})\mu
\psi_{\mu,\tau}\right.\nonumber\\
&&\qquad\qquad+\left.
(\mu-2\delta {\cal N}_{\vartheta}(\mu,\tau)^{-1})
\psi_{\mu-4\delta {\cal N}_{\vartheta}(\mu,\tau)^{-1},\tau}\right)\nonumber\\
&&+ C_-(\mu,\tau)
\left(\psi_{\mu-2\delta {\cal N}_{\vartheta}(\mu,\tau)^{-1},
\tau-2\delta {\cal N}_x(\mu,\tau)^{-1}}-
\psi_{\mu+2\delta {\cal N}_{\vartheta}(\mu,\tau)^{-1},
\tau-2\delta {\cal N}_x(\mu,\tau)^{-1}}\right)\nonumber\\
&=& 0\, .
\end{eqnarray}
with
\begin{eqnarray} \label{coeff}
 C_{\pm}(\mu,\tau) &=& 2\delta {\cal N}_{\vartheta}(\mu,\tau)^{-1}
(\sqrt{|\tau\pm 2\delta {\cal N}_x(\mu,\tau)^{-1}|}+ \sqrt{|\tau|}) \\
 C_0(\mu,\tau) &=& \sqrt{|\tau+\delta {\cal
N}_x(\mu,\tau)^{-1}|}-\sqrt{|\tau-\delta {\cal
N}_x(\mu,\tau)^{-1}|}\,.
\end{eqnarray}
(A total factor ${\cal N}_x{\cal N}_{\vartheta}^2$ for the number of
vertices drops out because the right hand side is zero in vacuum, but
would multiply the left hand side in the presence of a matter term.)

\section{Specific refinement models}

For further analysis one has to make additional assumptions on how
exactly the lattice spacing is changing with changing scales $\mu$ and
$\tau$. To fix this in general, one would have to use a full
Hamiltonian constraint and determine how its action balances the
creation of new vertices with increasing volume. Instead of doing
this, we will focus here on two geometrically motivated
cases. Technically simplest is a quantization where the number of
vertices in a given direction is proportional to the geometrical area
of a transversal surface. Moreover, the appearance of transversal
surface areas is suggested by the action of the full Hamiltonian
constraint which, when acting with an edge holonomy, creates a new
vertex along this edge (changing ${\cal N}_I$ for this direction) and
changes the spin of the edge (changing the area of a transversal
surface). It also agrees with \cite{APSII,BianchiImubar}, although the
motivation in those papers, proposing to use geometrical areas rather
than coordinate areas ${\cal A}_{IJ}$ in (\ref{loop}), is different.

Geometrically more intuitive is the case where the number of vertices
in a given direction is proportional to the geometrical extension of
this direction.\footnote{This behavior is introduced independently in
\cite{SchwarzNHol} where ``effective'' equations, obtained by
replacing connection components in the classical constraint by sines
and cosines of such components according to how they occur in the
quantized constraint, are analyzed for the Schwarzschild interior. The
results are complementary to and compatible with our stability
analysis of the corresponding difference equations below. We thank
Kevin Vandersloot for discussions on this issue.}  The resulting
difference equation will be more difficult to deal with due to its
non-constant step-size, but naturally gives rise to Misner-type
variables. This case will also be seen to have improved stability
properties compared to the first one using areas. In both cases,
${\cal N}\propto V$ is assumed, i.e.\ the lattice size increases
proportionally to volume. This is not necessary in general, and we
choose these two cases mainly for illustrative purposes. In fact,
constant ${\cal N}$ as in \cite{Bohr} and ${\cal N}\propto V$ first
used in \cite{APSII} are two limiting cases from the full point of
view, the first one without creating new vertices and the second one
without changing spin labels along edges since local lattice volumes
$V/{\cal N}$ remain constant. In general, both spin changes and the
creation of new vertices happen when acting with a Hamiltonian
constraint operator. Thus, one expects ${\cal N}\propto V^{\alpha}$
with some $0<\alpha<1$ to be determined by a detailed analysis of the
full constraint and its reduction to a homogeneous model. Even
assuming a certain behavior of ${\cal N}(V)$ without analyzing the
relation to a full constraint leaves a large field to be explored,
which can give valuable consistency checks. We will not do this
systematically in this paper but rather discuss a mathematical issue
that arises in any such case: initially, one has to deal with
difference equations of non-constant step-size which can be treated
either directly or by tranforming a non-equidistant difference
equation to an equidistant one. We first illustrate this for ordinary
difference equations since partial ones, as they arise in anisotropic
models, can often be reduced to this case.

\subsection{Ordinary difference equations of varying step-size}
\label{OrdEqual}

Let us assume that we have an ordinary difference equation for a
function $\psi_{\mu}$, which appears in the equation with
$\mu$-dependent increments $\psi_{\mu+\delta {\cal N}_1(\mu)^{-1}}$. To
transform this to a fixed step-size, we introduce a new variable
$\tilde{\mu}(\mu)$ such that $\tilde{\mu}(\mu+\delta/{\cal N}_1(\mu))=
\tilde{\mu}(\mu)+\delta \tilde{\mu}'/{\cal N}_1(\mu)+O(\delta^2)$ has a
constant linear term in $\delta$. (For the isotropic equation, ${\cal
N}_1$ is the vertex number only in one direction. The total number of
vertices in a 3-dimensional lattice is given by ${\cal N}={\cal N}_1^3$.)
 This is obviously satisfied if we
choose $\tilde{\mu}(\mu):= \int^{\mu}{\cal N}_1(\nu)\md\nu$. We then have
\begin{eqnarray}
 \psi_{\mu+\delta/{\cal N}_1(\mu)} &=&
 \tilde{\psi}_{\tilde{\mu}(\mu+\delta/{\cal N}_1(\mu))}=
\tilde{\psi}_{\tilde{\mu}+\delta+ \sum_{i=2}^{\infty}\frac{1}{i!}\delta^i
{\cal N}_1^{(i-1)}/{\cal N}_1^i}\\
&=& \tilde{\psi}_{\tilde{\mu}+\delta}+
\frac{1}{2}\delta^2\frac{{\cal N}_1'}{{\cal N}_1^2}\tilde{\psi}'+O(\delta^3)
\end{eqnarray}
where ${\cal N}_1^{(i)}$ denotes the $i$-th derivative of ${\cal
N}_1$.  Thus, up to terms of order at least $\delta^2$ the new
equation will be of constant step-size for the function
$\tilde{\psi}_{\tilde{\mu}}:=\psi_{\mu(\tilde{\mu})}$. (The derivative
$\tilde{\psi}'$ by $\tilde{\mu}$ may not be defined for any solution
to the difference equation. We write it in this form since such terms
will be discussed below in the context of a continuum or semiclassical
limit where derivatives would exist.)

It is easy to see that, for refining lattices, the additional terms
containing derivatives of the wave function are of higher order in
$\hbar$ and thus correspond to quantum corrections. For ${\cal
N}_1(\mu)\propto \mu^q$ as a positive power of $\mu$, which is the
expected case from lattice refinements related to the increase in
volume, we have
\[ 
 \frac{{\cal N}_1'}{{\cal N}_1^2}=\frac{q}{\mu {\cal N}_1(\mu)}= 
q\left(\frac{4\pi\gamma\lP^2}{3p}\right)^{1+q}
\]
relating $\mu$ to an isotropic triad component
$p=4\pi\gamma\lP^2\mu/3$ as it occurs in isotropic loop quantum
gravity \cite{IsoCosmo}. Moreover,
\[
 \tilde{\psi}'=\frac{\md\tilde{\psi}}{\md\tilde{\mu}}=
\frac{\md\mu}{\md\tilde{\mu}} \frac{\md\psi}{\md\mu}= \frac{1}{{\cal N}_1(\mu)}
\frac{\md\psi}{\md\mu} = -\frac{i}{2}
\frac{1}{{\cal N}_1(\mu)}\hat{c}\psi
\]
in terms of a curvature operator $\hat{c}=8\pi i\gamma G\hbar/3
\md/\md p= 2i\md/\md\mu$ which exists in a continuum
limit \cite{SemiClass}.  Thus,
\[
 \frac{{\cal N}_1'}{{\cal N}_1^2}\tilde{\psi}'\propto
 \left(\frac{\hbar}{p}\right)^{1+2q} \hat{c}\tilde{\psi}
\]
With $q$ positive (or just larger than $-1/2$) for a refining lattice,
there is a positive power of $\hbar$, showing that additional terms
arising in the transformation are quantum corrections.

This has two important implications. First, it shows that the correct
classical limit is obtained if lattices are indeed refined, rather
than coarsened, since $q$ is restricted for corrections to appear in
positive powers of $\hbar$. In anisotropic models, as we will see, the
behavior is more complicated due to the presence of several
independent variables. An analysis of the semiclassical limit can then
put strong restrictions on the behavior of lattices. Secondly, we can
implicitly define a factor ordering of the original constraint giving
rise to the non-equidistant difference equation by declaring that all
quantum correction terms arising in the transformation above should
cancel out with factor ordering terms. We then obtain a strictly
equidistant equation in the new variable $\tilde{\mu}$. For example, a
function ${\cal N}_1(\mu)=\sqrt{|\mu|}$ gives $\tilde{\mu}\propto
|\mu|^{3/2}$ such that the transformed difference equation will be
equidistant in volume rather than the densitized triad component.  For
this special case, factor orderings giving rise to a precisely
equidistant difference equation have been constructed explicitly in
\cite{APSII,BianchiImubar}.

\subsection{Number of vertices proportional to transversal area}

A simple difference equation results if the number of vertices is
proportional to the transversal area in any direction.\footnote{Since
this refers to the area, it is the case which agrees with the
motivation of \cite{APSII,BianchiImubar}.} In the $x$-direction we
have transversal surfaces given by symmetry orbits of area $p_c$,
using the line element (\ref{metric}), and thus
${\cal N}_x\propto\tau$. Transversal surfaces for an angular direction are
spanned by the $x$- and one angular direction whose area is $p_b$,
giving ${\cal N}_{\vartheta}\propto\mu$. Each minisuperspace direction has a
step-size which is not constant but independent of the other
dimension. Moreover, due to the simple form one can transform the
equation to constant step-size by using independent variables $\tau^2$
and $\mu^2$ instead of $\tau$ and $\mu$. Illustrating the general
procedure given before, a function $\tilde{\psi}_{\tau^2,\mu^2}$ acquires
constant shifts under the basic steps,
\begin{eqnarray*}
 \tilde{\psi}_{(\tau+n\delta/\tau)^2,(\mu+m\delta/\mu)^2}&=&
\tilde{\psi}_{\tau^2+2n\delta+n^2\delta^2/\tau^2,
\mu^2+2m\delta+m^2\delta^2/\mu^2}\\
&=& \tilde{\psi}_{\tau^2+2n\delta,\mu^2+2m\delta}+O(\tau^{-2})+O(\mu^{-2})
\end{eqnarray*}
up to terms which can be ignored for large $\tau$ and $\mu$. This is
sufficient for a straightforward analysis in asymptotic
regimes. Moreover, higher order terms in the above equation come with
higher derivatives of the wave function in the form
\[
 \frac{\tilde{\psi}'}{\tau^2}=
\frac{\gamma^2\lP^4}{p_c^2}\tilde{\psi}'=
-i\frac{(\gamma\lP^2)^3}{4p_c^3}\hat{c}\tilde{\psi}
\]
since $q=1$ compared to the discussion in Sec.~\ref{OrdEqual}. Due to
the extra factors of $\hbar$ (or even higher powers in further terms
in the Taylor expansion) any additional term adding to the constant
shift of $\tilde{\psi}_{\tau^2,\mu^2}$ can be attributed to quantum
corrections in a semiclassical limit. Accordingly, such terms can be
avoided altogether by a judicious choice of the initial factor
ordering of operators.

\subsection{Number of vertices proportional to extension}

Geometrically more intuitive, and as we will see below dynamically
more stable, 
is the case in which the number of vertices in each
direction is proportional to the extension of that direction measured
with the triad itself. This gives ${\cal N}_{\vartheta}\propto
\sqrt{|\tau|}$ and ${\cal N}_x\propto\mu/\sqrt{|\tau|}$, using the
classical co-triad (\ref{cotriadclass}). (One need not worry about the
inverse $\tau$ since the effective treatment of lattice refinements
pursued here is not valid close to a classical singularity where an
already small lattice with a few vertices changes. Singularities in
general can only be discussed by a direct analysis of the resulting
difference operators. Since only a few recurrence steps are necessary
to probe the scheme around a classical singularity, equidistant
difference operators are not essential in this regime. They are more
useful in semiclassical regimes where one aims to probe long evolution
times as in the examples below. Similar remarks apply to the horizon
at $\mu=0$ which, although a classical region for large mass
parameters, presents a boundary to the homogeneous model used for the
Schwarzschild interior.)  The behavior is thus more complicated than
in the first case since the step size of any of the two independent
variables depends on the other variable, too. First, it is easy to
see, as before with quadratic variables, that the volume label
$\omega=\mu\sqrt{|\tau|}$ changes (approximately) equidistantly with
each iteration step which is not equidistant for the basic variables
$\mu$ and $\tau$. But it is impossible to find a second, independent
quantity which does so, too. In fact, such a quantity $f(\mu,\tau)$
would have to solve two partial differential equations in order to
ensure that
\[
 f(\mu+n\delta {\cal N}_{\vartheta}(\mu,\tau)^{-1},\tau+m\delta
 {\cal N}_x(\mu,\tau)^{-1})\sim f(\mu,\tau)+n\delta
 {\cal N}_{\vartheta}(\mu,\tau)^{-1}\partial_{\mu}f(\mu,\tau)+ m\delta
 {\cal N}_x(\mu,\tau)^{-1}\partial_{\tau}f(\mu,\tau)
\]
changes only by a constant independent of $\tau$ and $\mu$. This
implies $\partial_{\mu}f(\mu,\tau)\propto\sqrt{|\tau|}$ and
$\partial_{\tau}f(\mu,\tau)\propto \mu/\sqrt{|\tau|}$ whose only
solution is $f({\mu,\tau})\propto\mu\sqrt{|\tau|}$ which is the volume
$\omega$.

We thus have to deal with non-equidistant partial difference equations
in this case which in general can be complicated. A possible
procedure to avoid this is to split the iteration in two steps since
an ordinary difference equation can always be made equidistant as
above (cancelling quantum corrections by re-ordering). We first
transform $\tau$ to the volume variable $\omega$ which gives, up to quantum
corrections, constant iteration steps for this variable. With the second
variable still present, a higher order difference equation
\begin{eqnarray}
&& C_0(\mu,\omega^2/\mu^2)(1+2\delta/\omega)\mu
\psi_{\mu(1+4\delta/\omega),\omega+4\delta}+ C_+(\mu,\omega^2/\mu^2)
\psi_{\mu(1+2\delta/\omega),\omega+3\delta}\nonumber\\
&& -
C_-(\mu,\omega^2/\mu^2)\psi_{\mu(1+2\delta/\omega),\omega+\delta}-
2C_0(\mu,\omega^2/\mu^2)(1+2\gamma^2\delta^2\mu^2/\omega^2)\mu
\psi_{\mu,\omega}\nonumber\\
&&-
C_+(\mu,\omega^2/\mu^2)\psi_{\mu(1-2\delta/\omega),\omega-\delta}+
C_-(\mu,\omega^2/\mu^2)\psi_{\mu(1-2\delta/\omega),\omega-3\delta}\nonumber\\
&&+C_0(\mu,\omega^2/\mu^2)(1-2\delta/\omega)\mu
\psi_{\mu(1-4\delta/\omega),\omega-4\delta} =0
\end{eqnarray}
results with
\begin{eqnarray}
 C_0(\mu,\omega^2/\mu^2) &=& \frac{\omega}{\mu}\left(\sqrt{1+
\frac{\delta}{\omega}}- \sqrt{1-\frac{\delta}{\omega}}\right)\\
 C_{\pm}(\mu,\omega^2/\mu^2) &=& 2\delta \left(1+\sqrt{1\pm
\frac{2\delta}{\omega}}\right)
\end{eqnarray}
derived from the original coefficients (\ref{coeff}).
The structure of this difference equation is quite different from the
original one: not only is it of higher order, but now only one value
of the wave function appears at each level of $\omega$, rather than
combinations of values at different values of $\mu$. Note also that
only the coefficient of the unshifted $\psi_{\mu,\omega}$ depends on
$\mu$.  This form of the difference equation is, however, a
consequence of the additional rotational symmetry and is not realized
in this form for fully anisotropic Bianchi models as we will see
below.

Proceeding with this specific case, we have to look at wave functions
evaluated at shifted positions $\mu(1+m\delta/\omega)$ with integer
$m$. At fixed $\omega=\omega_0$, we are thus evaluating the wave
function at values of $\mu$ multiplied with a constant, instead of
being shifted by a constant as in an equidistant difference
equation. This suggests to use the logarithm of $\mu$ instead of $\mu$
itself as an independent variable, which is indeed the result of the
general procedure.  After having transformed from $\tau$ to $\omega$
already, we have to use $\tau$ as a function of $\mu$ and $\omega$ in
the vertex number ${\cal N}_{\vartheta}$, which is $\tau(\mu,\omega)=
(\omega/\mu)^2$ after using $\omega=\mu\sqrt{\tau}$. Thus, ${\cal
N}_{\vartheta}(\mu,\tau(\mu,\omega))=
\sqrt{\tau(\mu,\omega)}=\omega/\mu$ now is not a positive power of the
independent variable $\mu$ and we will have to be more careful in the
interpretation of correction terms after performing the
transformation. (The lattice is coarsened with increasing anisotropy
at constant volume.) Naively applying the results of
Sec.~\ref{OrdEqual} to $q=-1$ would suggest that corrections come with
inverse powers of $\hbar$ which would certainly be damaging for the
correct classical limit. However, the factors change due to the
presence of the additional variable $\omega_0$ even though it is
treated as a constant. We have ${\cal N}_{\vartheta}'/{\cal
N}_{\vartheta}^2=-1/\omega_0= -(\gamma \lP^2/2)^{3/2}/V_0$ in terms of the
dimensionful volume $V$, while it would just be a constant $-1$
without the presence of $\omega$. The additional factor of
$\hbar^{3/2}$ ensures that corrections come with positive powers of
$\hbar$ for the correct classical limit to be realized.

For any $\omega_0$, we thus transform
$\tilde{\psi}_{\mu(1+m\delta/\omega_0)}$ to equidistant form by using
$\tilde{\tilde{\psi}}_{\tilde{\mu}}=
\tilde{\psi}_{\mu(\tilde{\mu})}$ with $\tilde{\mu}(\mu)=\log\mu$.
This transformation is possible since the second label $\omega_0$ is
now treated as a constant, rather than an independent variable of a
partial difference equation. (Recall that for the type of difference
equation discussed here there is only one variable, the volume, which
is equidistant under all of the original discrete steps.) Despite of
negative powers of some variables in the vertex numbers, we have the
correct classical limit in the presence of $\omega$.  As before, the
transformation is exact up to higher order terms which are quantum and
higher order curvature corrections. Defining the original constraint
operator ordering implicitly by the requirement that all those terms
are cancelled allows us to work with an equidistant difference
equation.

\subsection{Bianchi models}

As mentioned before, the transformed difference equation does not
become higher order for fully anisotropic Bianchi models. In this
case, we have three independent flux labels $\mu_I$, $I=1,2,3$, and
vertex numbers ${\cal N}_I$. Using vertex numbers proportional to the
spatial extensions for each direction gives ${\cal N}_1=\sqrt{\mu_2\mu_3/\mu_1}$, ${\cal
N}_2=\sqrt{\mu_1\mu_3/\mu_2}$ and ${\cal
N}_3=\sqrt{\mu_1\mu_2/\mu_3}$. As in the difference equation for the
Schwarzschild interior, the difference equation for Bianchi models
\cite{HomCosmo} uses values of the wave function of the form
$\psi_{\mu_1+2\delta/{\cal N}_1,\mu_2+2\delta/{\cal N}_2, \mu_3}$. One
can again see easily that the volume $\omega=\sqrt{|\mu_1\mu_2\mu_3|}$
behaves equidistantly under the increments,
\begin{eqnarray*}
 \omega(\mu_1+2\delta/{\cal N}_1,\mu_2+2\delta/{\cal N}_2, \mu_3)&=&
\sqrt{\left(\mu_1+2\delta\sqrt{\frac{\mu_1}{\mu_2\mu_3}}\right)
\left(\mu_2+2\delta\sqrt{\frac{\mu_2}{\mu_1\mu_3}}\right)\mu_3}\\
&=&
\sqrt{\mu_1\mu_2\mu_3+4\delta\sqrt{\mu_1\mu_2\mu_3}+4\delta^2}=
\omega+2\delta+O(\delta^2)\,.
\end{eqnarray*}
The leading order term of the difference equation in $\omega$ results
from a combination
\begin{eqnarray*}
 && C_1\psi_{\mu_1,\mu_2+2\delta/{\cal N}_2,\mu_3+2\delta/{\cal N}_3}+
C_2\psi_{\mu_1+2\delta/{\cal N}_1,\mu_2,\mu_3+2\delta/{\cal N}_3}+
C_3\psi_{\mu_1+2\delta/{\cal N}_1,\mu_2+2\delta/{\cal N}_2,\mu_3}\\
 &\approx& C_1\tilde{\psi}_{\mu_1,\mu_2+2\delta/{\cal N}_2,\omega+2\delta}+
C_2\tilde{\psi}_{\mu_1+2\delta/{\cal N}_1,\mu_2,\omega+2\delta}+
C_3\tilde{\psi}_{\mu_1+2\delta/{\cal N}_1,\mu_2+2\delta/{\cal
N}_2,\omega+2\delta}\\
&=&  C_1\tilde{\psi}_{\mu_1,\mu_2(1+2\delta/\omega),\omega+2\delta}+
C_2\tilde{\psi}_{\mu_1(1+2\delta/\omega),\mu_2,\omega+2\delta}+
C_3\tilde{\psi}_{\mu_1(1+2\delta/\omega),\mu_2(1+2\delta/\omega),\omega+2\delta}\\
&=:& \hat{C}_+ \tilde{\psi}_{\omega+2\delta}(\mu_1,\mu_2)
\end{eqnarray*}
where we used $1/{\cal N}_1= \sqrt{\mu_1/\mu_2\mu_3}= \mu_1/\omega$
and defined the operator $\hat{C}_+$ acting on the dependence of
$\psi$ on $\mu_1$ and $\mu_2$. Thus, unlike for the Schwarzschild
interior the difference equation does not become higher order in
$\omega$, and the highest order term does have a difference operator
coefficient in the remaining independent variables.

The recurrence proceeds as follows: 
We have a partial difference equation of the form
\[
 \hat{C}_+\tilde{\psi}_{\omega+2\delta}(\mu_1,\mu_2)+
 \hat{C}_0\tilde{\psi}_{\omega}(\mu_1,\mu_2)+ 
 \hat{C}_-\tilde{\psi}_{\omega-2\delta}(\mu_1,\mu_2)
\]
with difference  operators $\hat{C}_{\pm}$  and $\hat{C}_0$ acting  on
the dependence on $\mu_1$ and $\mu_2$.
In terms of initial data at two slices of $\omega$ we can solve
recursively for $\hat{C}_0\tilde{\psi}_{\omega}(\mu_1,\mu_2)+
\hat{C}_-\tilde{\psi}_{\omega-2\delta}(\mu_1,\mu_2)
=:\phi(\mu_1,\mu_2)$ and then, in each $\omega$-step, use boundary
conditions to solve the ordinary difference equation
\[
 \hat{C}_+\tilde{\psi}_{\omega+2\delta}(\mu_1,\mu_2)=\phi(\mu_1,\mu_2)\,.
\]
Although the operator $\hat{C}_+$ itself is not equidistant, this
remaining ordinary difference equation can be transformed to an
equidistant one by transforming $\mu_1$ and $\mu_2$ as in
Sec.~\ref{OrdEqual} (using that $\omega$ is constant and fixed for
this equation at any recursion step). 
With $\mu_3(\mu_1,\mu_2,\omega)=
\omega^2/\mu_1\mu_2$, we have lattice spacings ${\cal
N}_1(\mu_1,\mu_2,\omega)=\omega/\mu_1$ and ${\cal
N}_2(\mu_1,\mu_2,\omega)=\omega/\mu_2$ in terms of $\omega$ which are
already independent of each other. The two remaining variables $\mu_1$
and $\mu_2$ are thus transformed to equidistant ones by taking their
logarithms as encountered before.

Note the resemblance of the new variables, volume and two logarithms a
metric components at constant volume, to Misner variables
\cite{Mixmaster}.
This observation may be of interest in comparisons with
Wheeler--DeWitt quantizations where Misner variables have often been
used, making the Wheeler--DeWitt equation hyperbolic.

\section{Application: Stability of the Schwarzschild interior}

Now that we have several possibilities for the lattice spacings, we
consider their effect on the solutions of the Hamiltonian
constraint. In particular, these solutions may have undesirable
properties reminiscent of numerical instabilities, as it was indeed
noticed for the original quantization of the Schwarzschild interior in
\cite{InstabLRS}. Also problems in the presence of a positive
cosmological constant, described in the introduction, are of this
type. Recall that when one wishes to solve an ordinary differential
equation, for example, there are various discrete schemes that ensure
errors do not propagate as the number of time steps increases. Here we
are in the opposite situation -- instead of having the freedom to pick
the discrete version of a continuous equation, the discrete equation
itself is what is fundamental. Thus, like a badly chosen numerical
recipe, some choices of the functions $N_\tau$ and $N_\vartheta$ in
the constraint equation may quickly lead to solutions that are out of
control, and increase without bound. To test for this, we will use a
von Neumann stability analysis~\cite{InstabLRS} on the possible
recursion relations. The essential idea is to treat one of the
relation parameters as an evolution parameter, and decompose the rest
in terms of orthogonal functions, representing ``spatial'' modes of
the solution. This will give rise to a matrix that defines the
evolution of the solution; if the matrix eigenvalues are greater than
unity for a particular mode, that mode is unstable. In particular, a
relation $\sum_{k = -M} ^M a_{n + k} \psi_{n+k} = 0$ is equivalent to
a vector equation of the form ${\vec v}_n = Q(n) {\vec v}_{n-1}$,
where the column vector ${\vec v}_n = (\psi_{n + M}, \psi_{n + M -1},
\cdots,
\psi_{n - M + 1})^T$. The evolution of an eigenvector ${\vec w}$ of
the matrix $Q(n)$ is given by ${\vec w}_n = \lambda_w {\vec
w}_{n-1}$. Thus, when the size of the corresponding eigenvalue
$|\lambda_w| > 1$, the values in the sequence associated to ${\vec w}$
will grow as well.

With this in mind, we consider the choices of $N_x$ and $N_\vartheta$
discussed previously, starting with the case $N_x = \tau$ and
$N_\vartheta = \mu$. In the large $\mu, \tau$ limit for this choice,
the coefficients of the Hamiltonian constraint become
\[
C_\pm (\mu, \tau) \sim \frac{4 \delta \sqrt{\tau}}{\mu}, \qquad
C_0 (\mu, \tau) \sim \frac{\delta}{\tau^{3/2}}.
\]
In the asymptotic limit, the coefficients of the $\psi_{\mu \pm
2\delta /\mu, \tau}$ and $\psi_{\mu, \tau}$ terms go to $C_0 (\mu,
\tau) \mu$. As we saw in Section 3.2, we can choose a different set of
variables in which the step sizes are constant (up to ordering of the
operators). Plugging these asymptotic values into the Hamiltonian
constraint, and changing variables to ${\tilde \mu} = \mu^2 / 2$ and
${\tilde \tau} = \tau^2 / 2$ gives
\[
4 {\tilde \tau} (\psi_{{\tilde \mu} + 2\delta, {\tilde \tau} +
2\delta} - \psi_{{\tilde \mu} - 2\delta, {\tilde \tau} + 2\delta} +
\psi_{{\tilde \mu} - 2\delta, {\tilde \tau} - 2\delta} - \psi_{{\tilde
\mu} + 2\delta, {\tilde \tau} - 2\delta}) + \tilde \mu ( \psi_{{\tilde
\mu} + 4\delta, {\tilde \tau}} - 2 \psi_{{\tilde \mu}, {\tilde \tau}}
+ \psi_{{\tilde \mu} - 4\delta, {\tilde \tau}}) = 0.
\]
Because all the step sizes now are constants depending on $\delta$, we
define new parameters $m, n$ such that ${\tilde \mu} = 2 m \delta$ and
${\tilde \tau} = 2 n \delta$. Using $m$ as our evolution parameter and
$n$ as the ``spatial'' direction, we decompose the sequence as $\psi_{2m
\delta, 2n\delta} = u_m \exp(i n \omega)$. With this new function, the
recursion relation is written as
\[
2i n (u_{n+1} - u_{n-1}) - (m \sin \theta) u_n = 0.
\]
This is equivalent to the vector equation
\begin{equation}
\biggl[
\begin{array}{c}
u_{n+1} \\
u_n \\
\end{array}
\biggr]
=
\biggl[
\begin{array}{cc}
- \frac{im}{2n} \sin \theta	& 1 \\
1					& 0 \\
\end{array}
\biggr]
\biggl[
\begin{array}{c}
u_n \\
u_{n-1} \\
\end{array}
\biggr]
=
Q (m, n)
\biggl[
\begin{array}{c}
u_n \\
u_{n-1} \\
\end{array}
\biggr].
\end{equation}
The eigenvalues of the matrix $Q$ are
\[
\lambda_\pm = \frac{-im\sin \theta \pm \sqrt{16n^2 - m^2 \sin^2 \theta}}{4n}.
\]
When the discriminant $16n^2 - m^2 \sin^2 \theta \ge 0$, then
$|\lambda| = 1$, and the solution is stable; however, there are
unstable modes when $16n^2 - m^2 \sin^2 \theta < 0$. The most unstable
mode corresponds to the choice $\sin \theta = 1$, giving instabilities
in terms of the original variables when $\mu > 2 \tau$.
In this regime, all solutions behave exponentially rather than oscillating.
This region includes parts of the classical solutions for the
Schwarzschild interior even for values of $\mu$ and $\tau$ for which
one expects classical behavior to be valid. The presence of
instabilities implies, irrespective of the physical inner product,
that quantum solutions in those regions cannot be wave packets
following the classical trajectory, and the correct classical limit is
not guaranteed for this quantization, which is analogous to that
introduced in \cite{APSII,BianchiImubar}.

The situation is different when we consider the choices $N_x =
\sqrt{|\tau|}$ and $N_\vartheta = \mu/\sqrt{|\tau|}$, where we will
find a lack of instability. There is no choice of variables that
allows us to asymptotically approach a constant spacing recursion
relation, because of the mixing of the $\mu$ and $\tau$ variables in
the step size functions. Thus, we will make the assumption that in the
large $\mu, \tau$ limit, the solution does not change much under step
sizes $\delta N_x ^{-1}$ and $\delta N_\vartheta ^{-1}$. To see how
this affects the resulting stability of the solutions, we will look at
a simpler example first. If we start with the Fibonacci relation
$R_\tau \equiv \psi_{\tau + 1} - \psi_\tau - \psi_{\tau - 1} = 0$,
then the two independent solutions are of the form $\psi_\tau =
\kappa^\tau$, where $\kappa$ is either the golden ratio $\phi = (1 +
\sqrt{5})/2$ or else $-\phi^{-1}$. Only the latter solution meets the
criterion for stability, since $|\phi| > 1$. When we change this
relation to
\begin{equation}
\label{NC}
{\tilde R}_\tau \equiv \psi_{\tau + 1/\tau^n} - \psi_\tau - \psi_{\tau
- 1/\tau^n} = 0,
\end{equation}
with $n \ne 1$, the situation changes -- only one of the two solutions
outlined above will solve the relation asymptotically. In particular,
when we examine the error ${\tilde R}_\tau$ we get when we plug
$\kappa^\tau$ into the altered relation (\ref{NC}), i.e.
\[
{\tilde R}_\tau = \kappa^\tau (\kappa^{1/\tau^n} - 1 - \kappa^{-1/\tau^n}),
\]
the error is proportional to $\psi_\tau$ itself. As $\tau \to \infty$,
therefore, the error for the $\kappa = \phi$ solution grows without
bound, while that of $\kappa = -\phi^{-1}$ goes to zero. Thus, we see
in this situation a relation between the stability and the asymptotic
behavior of a solution.

Returning to the Schwarzschild relation, in the large $\mu, \tau$
limit the coefficient functions of the recursion relation are to
leading order
\[
C_\pm (\mu, \tau) \sim 4 \delta, \qquad
C_0 (\mu, \tau) \sim \frac{\delta}{\mu}.
\]
In turn, the relation itself becomes
\begin{eqnarray*}
&\ & 4 (\psi_{\mu + 2\delta/\sqrt{\tau}, \tau +
2\delta\sqrt{\tau}/\mu} - \psi_{\mu - 2\delta/\sqrt{\tau}, \tau +
2\delta\sqrt{\tau}/\mu} - \psi_{\mu + 2\delta/\sqrt{\tau}, \tau -
2\delta\sqrt{\tau}/\mu} + \psi_{\mu - 2\delta/\sqrt{\tau}, \tau -
2\delta\sqrt{\tau}/\mu}) \\ &\ & + (\psi_{\mu + 4\delta/\sqrt{\tau},
\tau} - 2 \psi_{\mu, \tau} + \psi_{\mu - 4\delta/\sqrt{\tau}, \tau})=
0.
\end{eqnarray*}
{}From this point on, we assume that we have a solution to this relation
which does not vary greatly when, for example, $\mu$ is changed by
$\pm 2\delta/\sqrt{\mu}$, and similarly for $\tau$. Both $N_x$ and
$N_\vartheta$ are constant to first order in shifts $\mu \pm 2 \delta
N_x ^{-1}$ and similarly for $\tau$, in the asymptotic limit. Thus, we
assume that $\alpha = 2 \delta N_x ^{-1}$ and $\beta = 2 \delta
N_\vartheta ^{-1}$ are constants, and use the scalings $\mu = \alpha
m$ and $\tau = \beta n$. When this is done, we get an equation similar
to the case when $N_x = \tau$ and $N_\vartheta = \mu$, but with
constant coefficients; this is the crucial difference that allows
stable solutions to the case here. Using the decomposition
$\psi_{\alpha m, \beta n} = u_n \exp(i m \theta)$, we arrive at the
matrix equation
\begin{equation}
\biggl[
\begin{array}{c}
u_{n+1} \\
u_n \\
\end{array}
\biggr]
=
\biggl[
\begin{array}{cc}
- \frac{i}{2} \sin \theta	& 1 \\
1					& 0 \\
\end{array}
\biggr]
\biggl[
\begin{array}{c}
u_n \\
u_{n-1} \\
\end{array}
\biggr].
\end{equation}
The matrix here has eigenvalues $\lambda$ with $|\lambda| = 1$ for all
$m, n$, so the solution is stable.
Using arguments as in the Fibonacci example, the non-equidistant
equation of the second scheme is shown to be stable.

\section{Conclusions}

Following \cite{InhomLattice}, we explicitly introduced loop quantum
cosmological models which take into account the full lattice structure
of inhomogeneous states. Such lattices are in general refined by
adding new vertices when acting with the Hamiltonian constraint. Thus,
also dynamical equations even in homogeneous models should respect
this property. Several interesting features arose: One obtains
non-equidistant difference equations which, when imposed for functions
on the whole real line as in isotropic loop quantum cosmology, are
more restrictive than equidistant ones due to the absence of
superselected sectors. This leaves the singularity issue unchanged
since for this one only needs to consider a few steps in the
equation. But a stability analysis of solutions and the verification
of the correct classical limit in all semiclassical regimes can be
more challenging. We presented an example for such an analysis, but
also introduced a procedure by which one can transform the resulting
equations to equidistant ones up to quantum corrections, which is
sufficient for a semiclassical analysis. Interestingly, properties of
the transformation itself provide hints to the correct semiclassical
behavior. As a side-result, we demonstrated that one particular
version of lattice refinements naturally gives rise to Misner-type
variables.

It is our understanding that this general procedure of defining
lattice refining models mostly agrees with the intuition used
specifically in isotropic models in \cite{APSII}, and adapted to
anisotropic ones in \cite{BianchiImubar}.\footnote{We thank A.\
Ashtekar for discussions of this point.} However, there are some
departures from what is assumed in \cite{APSII}. First, we do not see
indications to refer to the area operator while the area spectrum was
not only used in \cite{APSII} to fix the constant $\delta$ and the
volume dependence of the step size but in fact provided the main
motivation. Secondly, due to this motivation \cite{APSII} presents a
more narrow focus which from our viewpoint corresponds to only one
single refinement model. It has a vertex number proportional to
volume, which is a limiting case not realized by known full
Hamiltonian constraints, and puts special emphasis on geometrical
areas to determine the vertex number. Finally, commutators for inverse
volume operators are to be treated differently from \cite{APSII},
taking into account a lattice refining model which would not be
possible in a purely homogeneous formulation. As shown in the
appendix, this enlarges expected quantum corrections to the classical
functions.

We have discussed similar cases for illustration here, but keep a more
general viewpoint on the refinement as a function of volume. A
preliminary stability analysis for the Schwarzschild interior,
consistent with \cite{SchwarzNHol} indeed suggests that a behavior
different from what is suggested in \cite{APSII} is preferred, which
indicates that models can provide tight conditions for the general
analysis of quantum dynamics.  We emphasize that stability arguments
as used here are independent of physical inner product issues since
they refer to properties of {\em general solutions}.  A general
analysis as started here allows detailed tests of the full dynamics in
manageable settings, which can verify the self-consistency of the
framework of loop quantum gravity --- or possibly point to limitations
which need to be better understood.

\section*{Acknowledgements}

We thank Kevin Vandersloot for discussions. This work was supported in
part by NSF grant PHY0554771. GK is grateful for research support from
the University of Massachusetts and Glaser Trust of New York.

\begin{appendix}
\section{Inverse volume terms in homogeneous models and lattice
refinement}

We have seen in this paper, following \cite{InhomLattice}, that
Hamiltonian constraint operators with triad-dependent parameters in
holonomies allow one to model lattice refinements faithfully, with
interesting results and some improvements over the original
non-refining models. However, as always there are also some features
of inhomogeneous states and operators which are not present and
difficult to mimic in homogeneous models. Thus, even models
generalized in this way by allowing for lattice refinement effects
have to be interpreted with great care. While qualitative effects can
be investigated fruitfully to test the full framework, there is no
basis for drawing quantitative conclusions. The prime example is that
of commutator terms which appear ubiquitously in composite operators
of loop quantum gravity, such as the coefficients of difference
equations or also matter Hamiltonians.

In the main construction of this paper we used holonomies associated
with links of a lattice, rather than edges of a fixed coordinate
length. This allows us, effectively, to take into account lattice
refinements which change the number of vertices. It applies to
holonomies (\ref{loop}) along a closed loop used to quantize curvature
components which determine the step size of difference equations, but
also to the link holonomies used in commutators to quantize inverse
triad components based on (\ref{cotriad}). What is not modeled in
homogeneous models is the fact that a lattice operator makes use of
the {\em local} volume $\hat{V}_v$ at a given vertex $v$ where the
commutator is acting, rather than the {\em total} volume
$\hat{V}=\sum_v\hat{V}_v$ of the whole box in which the lattice is
embedded. In a fully inhomogeneous setting the difference does not
matter since volume contributions from vertices not touched by the
edge used in a commutator drop out in the end, $[h_e,\hat{V}]=
\sum_{v\in e} [h_e,\hat{V}_v]$. But in
homogeneous models there is a difference since volume contributions
from different vertices, in an exactly homogeneous setting, are all
identical. Thus, the total volume $V={\cal N}V_v$ is the number of
vertices multiplied with the local volume $V_v$. Then, $[h,\hat{V}_v]$
rather than $[h,\hat{V}]$ is expected as the contribution to
constraint operators from the inhomogeneous perspective. In
homogeneous models as in \cite{APSII}, on the other hand,
$[h,\hat{V}]$ is more straightforward to use. We now show that without
corrections this would imply crucial deviations from the inhomogeneous
behavior.

It is easy to see that commutators differ depending on whether the
local or total volume is used. For simplicity of the argument, we
proceed with an isotropic situation where $V=|p|^{3/2}$ in terms of
the basic isotropic densitized triad component $p$. A local lattice
flux, for a surface $S$ intersecting only a single link, would be
$\rho=\int_S\md^2y\tilde{p}=\ell_0^2p/L_0^2=p/{\cal N}^{2/3}$ for
links of coordinate length $\ell_0$, such that
$V_v=|\rho|^{3/2}=|p|^{3/2}/{\cal N}$ is the local volume. (We again
use a coordinate box of size $L_0^2$ and introduce rescaled flux
variables $p=L_0^2\tilde{p}$.) Isotropic states are spanned by
$e^{i\mu c/2}$ where $\mu\in{\mathbb R}$ is related to the triad
eigenvalues by $p_{\mu}=4\pi\gamma\lP^2\mu/3$.  Using a link holonomy
$h\sim e^{i\ell_0\tilde{c}/2}=e^{ic/2{\cal N}^{1/3}}$ which as a
multiplication operator increases $\mu$ by $1/{\cal N}^{1/3}$, a
commutator with the total volume will have eigenvalues of the form
\begin{equation} \label{commtotal}
 h^{-1}[h,\hat{V}]\sim V(\mu+1/{\cal N}^{1/3})-V(\mu-1/{\cal N}^{1/3})=
 |p+4\pi\gamma\lP^2/3{\cal N}^{1/3}|^{3/2}-|p-4\pi\gamma\lP^2/3{\cal
 N}^{1/3}|^{3/2}\,.
\end{equation}
If the local volume is used, on the other
hand, we have to refer to local edge labels
$\mu_e=3\rho/4\pi\gamma\lP^2$, rather than using the total $p$. Thus,
\begin{eqnarray}
 h_e^{-1}[h_e,\hat{V}_v]&\sim&
 V_v(\mu_e+1)-V_v(\mu_e-1)=|\rho+4\pi\gamma\lP^2/3|^{3/2}-
 |\rho-4\pi\gamma\lP^2/3|^{3/2}\nonumber \\ &=& {\cal
 N}^{-1}(|p+4\pi\gamma\lP^2{\cal N}^{2/3}/3|^{3/2}-
 |p-4\pi\gamma\lP^2{\cal N}^{2/3}/3|^{3/2})\,. \label{commlocal}
\end{eqnarray}
For large volume, $p\gg {\cal N}$, both expressions give the correct
classical limit $\frac{3}{2}{\cal N}^{-1/3}\sqrt{|p|}$ expected from
$\{e^{ic/2N^{1/3}},|p|^{3/2}\}$. However, quantum corrections, i.e.\
deviations from this classical limit for finite $p$, are much larger
for the second version using the local volume as it would occur in an
inhomogeneous quantization. The smooth classical function $\md V/\md
p$ in a Poisson bracket appears in discretized form by the large
step-size $N^{2/3}$ in (\ref{commlocal}) rather than the small one
$N^{-1/3}$ in (\ref{commtotal}). Perturbative corrections, derived by
Taylor expanding the difference terms and keeping higher order
corrections to the classical expression, are thus larger. (This can
have cosmological implications
\cite{InhomEvolve,HamPerturb,QuantCorrPert}.) Non-perturbative effects as
observed for the inverse scale factor operator in isotropic loop
quantum cosmology which has an upper bound at finite volume
\cite{InvScale}, start to arise for $p\sim {\cal N}^{2/3}$ when the local
volume is used but only at the much smaller $p\sim {\cal N}^{-1/3}$ for the
total volume. Since only the local volume is relevant for
inhomogeneous quantizations, quantum corrections from inverse volume
operators can be large.

Unfortunately, this effect is more difficult to mimic in exact
homogeneous models unlike the behavior of holonomies under lattice
refinements and has therefore been overlooked in \cite{APSII}. The
connection components appearing in holonomies can simply be divided by
a function ${\cal N}$ of triad components to implement shrinking edges
due to subdivision. This is not possible for the volume itself to use
a local version in a homogeneous model since, if we would divide the
total volume by the appropriate function of triad components, only a
constant would remain for an ${\cal N}$ proportional to volume and the
commutator would be zero. The only way to have this effect faithfully
implemented in a homogeneous model is to use higher SU(2)
representations for holonomies in commutators but {\em not} for
holonomies used in the loop to quantize curvature components. (This is
not possible if one writes the constraint as a single trace,
$\tr(h_{\alpha}h_e[h_e^{-1},\hat{V}])$ but can easily be done using
the equivalent form $\tr(\tau_ih_{\alpha})\tr(\tau_i
h_e[h_e^{-1},\hat{V}])$. We emphasize that higher representations for
commutators are advocated here only in exactly homogeneous models to
mimic inhomogeneous effects. Fully inhomogeneous operators usually
need not refer to higher representations.) In a representation of spin
$j$, matrix elements of holonomies contain exponentials
$\exp(imc/2{\cal N}^{1/3})$ with $-j\leq m\leq j$, which increases the
shifts in volume labels resulting from commutators. Resulting
expressions for commutators can be found in \cite{Ambig,ICGC}. If the
representation label $j$ is of the order ${\cal N}$, effects as they
result from lattice refinements and using the local volume are
correctly implemented. Accordingly, corrections from inverse triad
components quantized through commutators are much larger than they
would otherwise be.\footnote{This is the reason why inverse triad
corrections did not play a large role in recent numerical studies of
the so-called $\bar{\mu}$-quantization, as repeatedly emphasized in
that context \cite{APS,APSII,NegCurv,Inverse}. This is only the case
because this quantization is a hybrid version of lattice refinements
which are taken into account for holonomies but not for the local
volume used in commutators. Inverse triad corrections are thus
artificially, though subtly, suppressed in those quantizations,
departing from quantizations of inhomogeneous models.}

\end{appendix}


\end{document}